\documentclass[journal = jacsat]{achemso}

\usepackage{braket}
\usepackage{siunitx}
\usepackage{placeins}
\usepackage{amsmath}
\usepackage{amssymb}
\usepackage{cleveref}

\graphicspath{{Chain/}{VIB/}{UChain/}}

\SectionNumbersOn

\title{Efficient Steady-State Solver for the Hierarchical Equations of Motion Approach: Formulation and Application to Charge Transport through Nanosystems}

\author{C.\ Kaspar}
\affiliation{Institute of Physics, University of Freiburg, Hermann-Herder-Straße 3, D-79104 Freiburg, Germany
}
\email{christoph.kaspar@physik.uni-freiburg.de}

\author{M.\ Thoss}
\affiliation{Institute of Physics, University of Freiburg, Hermann-Herder-Straße 3, D-79104 Freiburg, Germany
}
\alsoaffiliation{EUCOR Centre for Quantum Science and Quantum Computing, University of Freiburg, Hermann-Herder-Straße 3, D-79104 Freiburg, Germany
}

\begin{document}

\maketitle

\begin{abstract}
An iterative approach is introduced, which allows the efficient solution of the hierarchical equations of motion (HEOM) for the steady-state of open quantum systems. 
The approach combines the method of matrix equations with an efficient preconditioning technique to reduce the numerical effort of solving the HEOM. 
Illustrative applications to simulate nonequilibrium charge transport in single-molecule junctions demonstrate the performance of the method.
\end{abstract}

\section{Introduction}
The hierarchical equations of motion (HEOM) formalism (also called hierarchical quantum master equation (HQME) method in the context of nonequilibrium electron transport) is an accurate and efficient approach to simulate the dynamics of open quantum systems. 
It was originally introduced by Tanimura and Kubo \cite{101143JPSJ58101,101143JPSJ75082001} to describe the relaxation dynamics of open quantum systems coupled to bosonic baths. 
Later, it was extended by Yan $et$ $al.$ \cite{doi10106312938087,Zheng2012,101103PhysRevLett109266403,Zheng2009,Zheng2013,Cheng2015,doi101002wcms1269} and H{\"a}rtle $et$ $al.$ \cite{PhysRevB88235426,PhysRevB90245426,101103PhysRevB92085430,PhysRevB94121303,PhysRevB94201407} to describe nonequilibrium dynamics and transport in nanosystems. For a recent overview of the method, extensions and applications, see Ref.\ \citenum{doi10106350011599} and references therein.

The HEOM approach is formulated as a density matrix scheme, which generalizes perturbative quantum master equation methods by including higher-order contributions and non-Markovian memory. 
It thus allows for the systematic convergence of the results, i.e., it is a numerically exact method.
This is achieved by introducing a set of hierarchically coupled auxiliary density operators, which are defined in the reduced system Hilbert space. However, accurate calculations with the HEOM method require significant numerical effort. 
Besides the total amount of emerging operators, their size, i.e., the dimension of the reduced system Hilbert space, mainly determines the computational cost of HEOM calculations. 
For example, systems at low temperatures with many system degrees of freedom require very time-consuming computations. 
Hence, various approaches have been developed to improve the efficiency of the HEOM formalism, such as fast propagation algorithms \cite{doiorg10106313077918,doi101021ct501066k}, scaling methods \cite{doiorg10106313077918,doi101021jp109559p}, the utilization of matrix product states \cite{doi10106315026753,doi10106315099416}, sophisticated ways to decompose the bath correlation functions \cite{doi10106313484491,Hu2011,doi10106314999027,doi10106315096945}, and hierarchy truncation schemes \cite{doi101143JPSJ743131,doi10106311850899,doi10106314914514}.
In addition, the implementation of high-performance algorithms on GPGPU architectures \cite{Kreisbeck2011,Kreisbeck2012,Hein2012,Kreisbeck2014,doi101021acsjctc5b00488,C6FD00088F}, single-node parallel computers \cite{doi101021ct3003833,Kreisbeck2014}, and distributed multicore clusters \cite{doiorg101002jcc25354} has been reported.

In this work, we outline an iterative algorithm to efficiently solve the HEOM for the steady-state. 
The approach employs the method of matrix equations,\cite{simoncini2016computational,datta2004numerical,qureshi1995lyapunov,103770jissn20952651201904008} which has been utilized before in the context of open quantum systems, e.g., to solve perturbative quantum master equations\cite{PhysRevB70205334} such as the Redfield equation \cite{doi1010631467222}.
Employed within the HEOM approach, the iterative algorithm in combination with an efficient preconditioning technique extends the applicability of the formalism to larger systems and more challenging regimes.
We demonstrate the performance of the method using illustrative applications to charge transport in molecular junctions.

The remainder of this paper is organized as follows. 
We introduce the theoretical methodology in \cref{Theory}, including a description of the examined model system in \cref{Model} and a brief review of the HEOM formalism in \cref{HEOM}. 
The iterative steady-state solver is outlined in \cref{Iterative}, where we introduce the mathematical framework in \cref{GCME} and present improvements to increase the computational efficiency in \cref{PrecondSec,Trunc}. 
In \cref{Results}, the numerical performance of the iterative approach is illustrated by applications to representative model systems, and the results are compared to those from previously applied conventional steady-state methods. 
We close with conclusions in \cref{Conclusion}.

\section{Theory} \label{Theory}
In this section, we recapitulate the formulation of the HEOM method for a nonequilibrium open quantum system. 
To be specific, we consider the example of charge transport in a molecular junction. 
It is noted, however, that the steady-state solver introduced in this paper can also be used for any other application of the HEOM method.

\subsection{Model System} \label{Model}
As an example of a nonequilibrium open quantum system, we consider a molecular junction consisting of a molecule coupled to two macroscopic electrodes (leads), which represent the environments. 
The Hamiltonian of the composite system is given by
\begin{align} 
 H          = & H^{\phantom{\dagger}}_{\rm S} + H^{\phantom{\dagger}}_{\rm B} + H^{\phantom{\dagger}}_{\rm SB},\label{totalH} \\
 H^{\phantom{\dagger}}_{\rm B}  = &\sum_{k \in \rm L/R} \epsilon^{\phantom{\dagger}}_k c^{\dagger}_k c^{\phantom{\dagger}}_k, \\
 H^{\phantom{\dagger}}_{\rm SB} = &\sum_{k \in {\rm L/R} ,m} (V^{\phantom{\dagger}}_{k,m} c^{\dagger}_k d^{\phantom{\dagger}}_m + V^{*}_{k,m}d^{\dagger}_m c^{\phantom{\dagger}}_k) \label{tunnelingH}.
\end{align}
The molecule, which in the following is referred to as the system, is described by the Hamiltonian $H_{\rm S}$ which can have a general form and may, for example, include electron-electron interactions or electronic-vibrational coupling. 
The two electrodes are characterized by the noninteracting Hamiltonian $H_{\rm B}$ and involve a continuum of electronic states with energy $\epsilon_k $.
Lead electrons are annihilated/created by the operators $c^{\phantom{\dagger}}_k/c^{\dagger}_k$. 
The interactions between the Fermionic environments and the system are summarized in $H_{\rm SB}$. 
The molecular electronic state $m$ with corresponding annihilation/creation operators $d^{\phantom{\dagger}}_m/d^{\dagger}_m$ couples to the states in the leads with the tunneling matrix element $V_{k,m}$. 
The coupling is determined by the level-width function of lead $K \in {\rm L/R}$
\begin{align}
 \Gamma^{\phantom{\dagger}}_{Kmn}(\epsilon)=2 \pi \sum_{k \in {\rm L/R}} V^{\phantom{*}}_{k,m} V^{*}_{k,n} \delta(\epsilon - \epsilon^{\phantom{*}}_k).
\end{align}

The numerical complexity of the proposed iterative solving technique can be considerably reduced by transforming the total Hamiltonian in Eq.\ \eqref{totalH} to the eigenbasis of the molecular Hamiltonian $H_{\rm S}$ using the unitary transformation
\begin{align}
\tilde{H} = & U^{\dagger} H U = \tilde{H}^{\phantom{\dagger}}_{\rm S} + H^{\phantom{\dagger}}_{\rm B} + \tilde{H}^{\phantom{\dagger}}_{\rm SB}.
\end{align}
As discussed below, this enables an efficient strategy\cite{PhysRevB94201407,doi10106314914514} to truncate the HEOM.

To describe the system-bath interaction within the HEOM approach, we introduce the bath coupling operators employing a bath-interaction picture
\begin{align}
 F^{\sigma}_{Km} (t) = e^{i H_{\rm B} t} \big( \sum_{k \in {\rm L/R}} V^{\phantom{\sigma}}_{k,m} c^{\sigma}_k \big) e^{-i H_{\rm B} t},
\end{align}
with $\sigma=\pm$ and $c^{-(+)}_k \equiv c^{(\dagger)}_k$. For system-bath couplings characterized by $H_{\rm SB}$ in Eq.\ \eqref{tunnelingH}, the effect of the environment on the molecule can be entirely described by the two-time correlation functions of the free bath (we set $\hbar =k_{\rm B}= e = 1$ throughout)
\begin{align}
 C^{\sigma}_{Kmn}(t-\tau) = & \braket{ F^{\sigma}_{Km}(t) F^{\bar{\sigma}}_{Kn}(\tau) }_{\rm{B}},
\end{align}
with $\bar{\sigma} \equiv -\sigma$. 
The central step to obtain a closed set of equations within the HEOM framework is to expand the correlation functions by a series of exponential functions \cite{doi10106312938087,Zheng2012}
\begin{align} \label{Correlationfunctionapprox}
 C^{\sigma}_{Kmn}(t-\tau) \approx V^{\phantom{\sigma}}_{K,m} V^{\phantom{\sigma}}_{K,n} \sum_{l=0}^{N_{\rm P}} \eta^{\phantom{\sigma}}_{K,\sigma,l} e^{-\gamma_{K,\sigma,l} (t-\tau)}.
\end{align}
Various sum-over poles schemes have been used to determine the coefficients $\eta^{\phantom{\sigma}}_{K,\sigma,l}$ and exponents $\gamma^{\phantom{\sigma}}_{K,\sigma,l}$, such as the low-frequency logarithmic discretization \cite{doi10106314999027}, the Fano spectrum decomposition scheme \cite{doi10106315096945}, and the Pad\'{e} approximation \cite{doiorg10106313077918,doi10106313484491,Hu2011}. 
Due to its superior efficiency \cite{doi10106314914514}, we employ the latter method in this work. 
For numerical applications, the expansion in Eq.\ \eqref{Correlationfunctionapprox} is truncated at $N_{\rm P}$ exponential terms, which we refer to as the number of Pad\'{e} poles in the following sections. 
Moreover, we consider a level-width function given by a single Lorentzian
\begin{align}
 \Gamma^{\phantom{\dagger}}_{Kmn}(\epsilon) = 2 \pi \frac{V^{\phantom{\dagger}}_{K,m} V^{\phantom{\dagger}}_{K,n} W^2_K}{(\epsilon-\mu^{\phantom{\dagger}}_K)^2 + W^2_K},
\end{align}
with band-width $W_K$ and chemical potential $\mu_K$ of lead $K$. We note that an extension to realistic band structures was recently proposed.\cite{doi10106315041716}

\subsection{Hierarchical Equation of Motion Approach} \label{HEOM}
The HEOM formalism (also referred to as HQME method in this context) has been widely applied as an accurate and reliable method for studying charge transport in single-molecule junction scenarios.\cite{doi10106312938087,Zheng2012,101103PhysRevLett109266403,PhysRevB94201407,PhysRevB88235426,doi101002wcms1269}
We omit a detailed derivation and closely follow Refs.\ \citenum{doi10106312938087,PhysRevB88235426,PhysRevB94201407}. 
By introducing hierarchically coupled auxiliary density operators (ADOs), the HEOM method describes the nonequilibrium dynamics of the composite system.
The equation of motion for the $n$th tier ADO reads
\begin{align} \label{HEOMeq}
 \dot{\tilde{\rho}}^{(n)}_{j_n \ldots j_1} = & -i [\tilde{H}^{\phantom{\dagger}}_{\rm S},\tilde{\rho}^{(n)}_{j_n \ldots j_1}] - \sum^n_{k=1}\gamma^{\phantom{\dagger}}_{j_k}\tilde{\rho}^{(n)}_{j_n \ldots j_1} \nonumber \\
  &- i \sum_j \mathcal{\tilde{A}}^{\bar{\sigma}}_K \tilde{\rho}^{(n+1)}_{jj_n \ldots j_1} -i \sum^{n}_{k=1} \mathcal{\tilde{C}}^{\phantom{\dagger}}_{j_k} \tilde{\rho}^{(n-1)}_{j_n \ldots j_1/j_k}.
\end{align}
Here, $\tilde{\rho}^{(0)}$ denotes the reduced density operator and $\tilde{\rho}^{(n>0)}_{j_n \ldots j_1}$ are the ADOs, for which we have defined the multi-index $j=(K,\sigma,l)$. 
The multi-indices $jj_n \ldots j_1$ are the combination of the indices $j$ and $j_n \ldots j_1$, while in $j_n \ldots j_1/j_k$ the multi-index $j_k$ is omitted.
The ADOs are defined in the reduced system Hilbert space and contain information on bath-related observables such as the electrical current for lead $K$
\begin{align}
 I^{\phantom{\dagger}}_K=i \sum_{m,l} V^{\phantom{\dagger}}_{K,m} {\rm Tr}^{\phantom{\dagger}}_{\rm S} (d^{\phantom{\dagger}}_m U^{\dagger} \tilde{\rho}^{(1)}_{(K,+,l)} U - {\rm h.c.} ),
\end{align}
where ${\rm Tr}_{\rm S}$ denotes the trace over system degrees of freedom. 
The operator $\mathcal{\tilde{A}}$ ($\mathcal{\tilde{C}}$) is responsible for the tier-up (tier-down) coupling. 
The corresponding action on the ADOs is given by
\begin{subequations}
\begin{alignat}{2} \label{eq: HEOMOp}
  \mathcal{\tilde{A}}^{\bar{\sigma}}_K \tilde{\rho}^{(n+1)}_{jj_n\ldots j_1} = & \sum_m V^{\phantom{\dagger}}_{K,m} ( && U^{\dagger}d^{\bar{\sigma}}_m U \tilde{\rho}^{(n+1)}_{jj_n\ldots j_1} +  \nonumber \\
&  && (-)^{n} \tilde{\rho}^{(n+1)}_{jj_n\ldots j_1} U^{\dagger} d^{\bar{\sigma}}_m U ) ,\\
  \mathcal{\tilde{C}}^{\phantom{\dagger}}_j \tilde{\rho}^{(n-1)}_{j_n\ldots j_1/j} = & \sum_m V^{\phantom{\dagger}}_{K,m} ( && \eta^{\phantom{\dagger}}_j U^{\dagger}d^{\sigma}_m U \tilde{\rho}^{(n-1)}_{j_n\ldots j_1/j} +  \nonumber \\
 & && (-)^{n} \eta^{*}_j \tilde{\rho}^{(n-1)}_{j_n\ldots j_1/j} U^{\dagger} d^{\sigma}_m U ),
\end{alignat}
\end{subequations}
and is, in contrast to the slightly different formulation in Refs.\ \citenum{doi10106312938087,PhysRevB88235426}, modified by the unitary transformations $U$ and $U^{\dagger}$.
To reduce the amount of hierarchically coupled equations, we exploit the Hermitian conjugate relation of the ADOs that is introduced in Ref.\ \citenum{doi10106312938087},
\begin{align}
 \tilde{\rho}^{(n),\dagger}_{j_n \ldots j_1} = \tilde{\rho}^{(n)}_{\bar{j}_1 \ldots \bar{j}_n},
\end{align}
with $\bar{j} = (K,\bar{\sigma},l)$.

In principle, the HEOM defined in Eq.\ \eqref{HEOMeq} provides an exact description of open quantum systems. 
Recently, Han $et$ $al.$ demonstrated the exact truncation \cite{doi10106315034776} of the hierarchy at a finite level for Fermionic baths.
For converged results, the hierarchy can usually be terminated at a lower level $N_{\rm T}$, since the observables of interest, e.g., electronic populations, vibrational excitation, or the current through the molecular junction, are encoded in the reduced density operator or lower tier ADOs.
The maximal level of the hierarchy $N_{\rm T}$ determines the number of coupled equations $N_{\rm ADO}$ in the HEOM and thus the numerical effort. 
According to Ref.\ \citenum{doi10106313123526}, a rough estimate is given by
\begin{align} \label{eq: NADO}
 N_{\rm ADO} = \frac{1}{2} \sum^{N_{\rm T}}_{n=1} \frac{(n+M-1)!}{n!(M-1)!},
\end{align}
with $M=8(N_{\rm P}+1)$.
The challenge of the HEOM method is the rapid increase of the number of coupled equations, resulting in very time-consuming calculations, such as those in the scenario of strong molecule-lead coupling or low temperatures.
Additionally, for systems with a large reduced system Hilbert space dimension, the applicability of the HEOM method can be challenging.

\section{Iterative Steady-State Solver} \label{Iterative}
In nonequilibrium scenarios such as charge or heat transport, the steady-state is often of primary interest, which can be obtained in two ways.
One possibility is to propagate Eq.\ \eqref{HEOMeq} in time until the steady-state is reached.
To this end, a variety of real-time propagation algorithms has been applied, such as exponential integrators \cite{doi101021ct501066k,doi101021acsjctc5b00488} or classical \cite{Numericalrecipes,doiorg10106313077918} and adaptive time-step size \cite{doi101021ct3003833} Runge-Kutta methods. 
The major disadvantage of this strategy is the long propagation time that is often required for the system to relax to the steady-state, such as in the case of strong non-Markovian effects \cite{doiorg10106313077918}. 
According to our experience, even efficient implementations on GPGPU devices \cite{Kreisbeck2011,Kreisbeck2012,Hein2012,Kreisbeck2014,doi101021acsjctc5b00488,C6FD00088F} result in expensive computations.

As an alternative, directly solving for the steady-state may circumvent the issue of long relaxation times. \cite{PhysRevB94201407,PhysRevB97235429} 
This corresponds to solving Eq.\ \eqref{HEOMeq} with the left-hand side set to zero, i.e.,
\begin{align} \label{HEOM_ss}
 0 = & -i [\tilde{H}^{\phantom{\dagger}}_{\rm S},\tilde{\rho}^{(n)}_{j_n \ldots j_1}] - \sum^n_{k=1}\gamma^{\phantom{\dagger}}_{j_k}\tilde{\rho}^{(n)}_{j_n \ldots j_1} \nonumber \\
  &- i \sum_j \mathcal{\tilde{A}}^{\bar{\sigma}}_K \tilde{\rho}^{(n+1)}_{jj_n \ldots j_1} -i \sum^{n}_{k=1} \mathcal{\tilde{C}}^{\phantom{\dagger}}_{j_k} \tilde{\rho}^{(n-1)}_{j_n \ldots j_1/j_k}.
\end{align}
This strategy requires to set up and store a single coefficient matrix that represents the HEOM, which results in excessive memory usage.
In such scenarios, the methods of choice are iterative techniques, such as the biconjugate gradient (BICG) scheme \cite{doi101002wcms1269}, the transpose-free quasi-minimal residual (TFQMR) approach \cite{doi101002wcms1269}, the biconjugate gradient stabilized (BICGSTAB) algorithm \cite{doi101021ct3003833} and the self-consistent blocked Jacobi method \cite{doi10106314995424}.
The first two strategies utilize the sparsity of the HEOM \cite{doi10106314914514}; however, even storing only the nonzero elements of the coefficient matrix may result in a vast memory usage, as shown in Ref.\ \citenum{doi10106314914514}. 
As an alternative, the Jacobi method developed by Zhang $et$ $al.$ can be entirely defined in the reduced system Hilbert space. 
In this work, we generalize the idea by introducing general coupled matrix equations to utilize robust iterative methods and efficient tools as preconditioning or hierarchy truncation schemes.

\subsection{HEOM as General Coupled Matrix Equations} \label{GCME}
The basic idea of the iterative approach formulated in this work is the identification of the HEOM in Eq.\ \eqref{HEOM_ss} with general coupled matrix equations.\cite{simoncini2016computational,datta2004numerical,qureshi1995lyapunov,103770jissn20952651201904008}
This type of equation involves many smaller matrices instead of utilizing a single coefficient matrix with a large dimension.
In the context of open quantum systems, the idea has been previously applied to obtain the solution of perturbative quantum master equations,\cite{PhysRevB70205334,PhysRevB95085423} particularly the Redfield master equation\cite{doi1010631467222}.
To formulate the mapping to general coupled matrix equations for the HEOM method, we represent the hierarchical equations for the steady-state in Eq.\ \eqref{HEOM_ss} as $N_{\rm ADO} +1$ coupled matrix equations and define the corresponding linear operator
\begin{align}
 \mathcal{N}^{(n)}_{j_n \ldots j_1} (X) = &-i [\tilde{H}^{\phantom{\dagger}}_{\rm S},\tilde{\rho}^{(n)}_{j_n \ldots j_1}] - \sum^n_{k=1}\gamma^{\phantom{\dagger}}_{j_k}\tilde{\rho}^{(n)}_{j_n \ldots j_1} \nonumber \\
  &- i \sum_j \mathcal{\tilde{A}}^{\bar{\sigma}}_K \tilde{\rho}^{(n+1)}_{jj_n \ldots j_1} \nonumber \\
  & -i \sum^{n}_{k=1} \mathcal{\tilde{C}}^{\phantom{\dagger}}_{j_k} \tilde{\rho}^{(n-1)}_{j_n \ldots j_1/j_k},
\end{align} 
where $X$ is obtained by rearranging the reduced density operator and all ADOs
\begin{align}
 X&=(\tilde{\rho}^{(0)},\tilde{\rho}^{(1)}_{j_1},\tilde{\rho}^{(2)}_{j_2 j_1},\ldots). \label{eq: Xdef}
\end{align}
For a compact form, we introduce the linear operator $\mathcal{M}$
\begin{align}
X &\rightarrow \mathcal{M}(X) = 
\begin{pmatrix}
 \mathcal{N}^{(0)} (X) \\
 \mathcal{N}^{(1)}_{j_1} (X) \\
 \mathcal{N}^{(2)}_{j_2 j_1} (X)\\
 \vdots
\end{pmatrix}.
\end{align}
All numerical operations that occur are matrix-matrix multiplications whose computational effort scales with the third power of the reduced system Hilbert space dimension. 
The resulting coupled matrix equations can be recast in the form
\begin{align} \label{eq: HEOMMF}
 \mathcal{M}(X) = 0 = \dot{X},
\end{align}
which needs to be solved under the normalization constraint ${\rm Tr}(\tilde{\rho}^{(0)})=1$.
We note that the form of Eq.\ \eqref{eq: HEOMMF} crucially differs from those of ordinary linear equations.

Thus, the challenge is to iteratively solve the obtained general coupled matrix equations. 
In recent years, many authors have extended existing iterative methods originally introduced for ordinary linear equations to treat equations of the same type as Eq.\ \eqref{eq: HEOMMF}.\cite{ZHANG20119380,103770jissn20952651201904008,ZHOU2009327,HAJARIAN201437,LI20123545,Toutounianlsmr,BEIK20114605} 
The basic idea is to map the matrix equations onto corresponding linear equations using the Kronecker product and the vectorization operator.
Ordinary iterative methods for linear equations can then be utilized to solve for the solution, while the emerging matrix-vector multiplications are replaced by matrix-matrix operations.
Hence, the new approach inherits both the advantages and disadvantages of the original method.\cite{8528380} 
Since a unified convergence theory has not yet been developed for coupled matrix equations, the issues of iterative schemes such as breakdown, slow convergence rates, and stagnation need to be explained by the properties of the matrix representing the HEOM in Eq.\ \eqref{HEOMeq} in the reduced system Liouville space.

To identify an iterative technique for the HEOM method that does not suffer from convergence problems, we did extensive tests of several approaches, including BICGSTAB \cite{ZHANG20119380}, generalized product-type based on biconjugate gradient (GPBiCG\cite{6987397} and GPBiCG(m,l)\cite{103770jissn20952651201904008}), gradient-based iterative \cite{ZHOU2009327}, conjugate gradient squared (CGS\cite{HAJARIAN201437}), conjugate gradient on normal equations (CGNE\cite{ZHANG20119380}), and least-squares (LSQR\cite{LI20123545} and LSMR\cite{Toutounianlsmr}) algorithms. 
Our numerical experiments revealed that the BICGSTAB method tends to stagnate due to the well-known problem of large imaginary parts of the eigenvalues of the matrix in the reduced system Liouville space.\cite{Sleijpen1993BICGSTABL,doi1011370914062} 
To circumvent this issue, the GPBiCG(m,l) approach was developed on the basis of the BICGSTAB method. 
However, with this approach we also observed a significantly reduced convergence speed for increasing imaginary parts of the eigenvalues. 
Using the gradient-based iterative method, we found the slow convergence speed reported in Refs.\ \citenum{ZHANG20119380,HAJARIAN201437,101007s1163301407620,6987397}, while the CGS approach led to insufficiently accurate solutions, as described in Ref.\ \citenum{fokkema1996generalized}. 
On the other hand, our numerical experiments revealed that the CGNE, LSQR, and LSMR schemes have good and similar convergence properties and remain robust over a broad parameter regime. 
Based on these tests, we chose the CGNE algorithm outlined in Ref.\ \citenum{ZHANG20119380} as the best compromise between accuracy, required CPU-time, stability, and memory usage.
For this approach, the convergence rate is primarily determined by the squared condition number\cite{saad2003iterative} of the matrix defined in the Liouville space. 
An increased value of this quantity results in an increase in the number of iterative steps required.

\subsection{Preconditioning in the Reduced System Hilbert Space} \label{PrecondSec}
Iterative techniques may be affected by slow convergence rates or the lack of robustness. The key tool to enhance their efficiency is preconditioning.\cite{saad2003iterative,vandervorst_2003,axelsson_1994} 
The idea is to convert the original equations into a modified system with the same solution but numerically superior properties.
In the context of general coupled matrix equations, we define the action of a preconditioner as a linear operator $\mathcal{P}$ that transforms Eq.\ \eqref{eq: HEOMMF} into
\begin{align} 
 0 = & \mathcal{M} \Big(\mathcal{P}(Y)\Big),  \\
 Z = & \mathcal{P}(Y),\label{PrecondIntermediate}\\
 Y = & \mathcal{P}^{-1} (X). 
\end{align}
Here, $Y$ and $Z$ are defined equivalently to $X$ in Eq.\ \eqref{eq: Xdef}
\begin{subequations}
\begin{align}
 Y&=(\tilde{\rho}^{(0)}_{Y},\tilde{\rho}^{(1)}_{Y,j_1},\tilde{\rho}^{(2)}_{Y,j_2 j_1},\ldots), \\
 Z&=(\tilde{\rho}^{(0)}_{Z},\tilde{\rho}^{(1)}_{Z,j_1},\tilde{\rho}^{(2)}_{Z,j_2 j_1},\ldots), 
\end{align}
\end{subequations}
with the ADOs $\tilde{\rho}^{(n)}_{Y/Z,j_n \ldots j_1}$ required for intermediate storage.

A variety of different preconditioning methods exists, such as the incomplete LU factorization \cite{Meijerink1977,chen_2005,chow2015fine} or the sparse approximate inverse \cite{Grote1997,benzi1998sparse,chow1998approximate}. 
However, the application of these well established approaches to general coupled matrix equations is not straightforward since there is generally no representation of the involved matrices in the reduced system Hilbert space. 
Nevertheless, to use the advantages of this key technique we introduce the preconditioner
\begin{align} \label{Precond1}
 X \rightarrow \mathcal{P}(X) = & 
 \begin{pmatrix}
  \tilde{\rho}^{(0)} \\
  \mathcal{Q}^{(1)}_{j_1} (X) \\
  \mathcal{Q}^{(2)}_{j_2 j_1} (X) \\
  \vdots
 \end{pmatrix},
\end{align}
with
\begin{align} \label{Precond}
 \mathcal{Q}^{(n)}_{j_n \ldots j_1} (X) = &-i [\tilde{H}^{\phantom{\dagger}}_{\rm S},\tilde{\rho}^{(n)}_{j_n \ldots j_1}] - \sum^n_{k=1}\gamma^{\phantom{\dagger}}_{j_k}\tilde{\rho}^{(n)}_{j_n \ldots j_1}.
\end{align}
This form is motivated by the structure of the HEOM in the limit of vanishing coupling between different ADOs, i.e., neglecting the operators $\mathcal{\tilde{A}}$ and $\mathcal{\tilde{C}}$ in Eq.\ \eqref{HEOMeq}.
Using a more sophisticated preconditioner by incorporating this coupling leads to another, possibly smaller, general coupled matrix equation. 
However, the additional computational effort used to solve this matrix equation contradicts the important requirement \cite{saad2003iterative,vandervorst_2003} for $\mathcal{P}$ to inexpensively calculate $Y$ in Eq.\ \eqref{PrecondIntermediate} and prevents the efficient application of preconditioners other than that defined in Eqs.\ \eqref{Precond1} and \eqref{Precond}.

The remaining task is to solve for $Y$ in Eq.\ \eqref{PrecondIntermediate}, i.e., to compute the inverse action of the linear operator $\mathcal{P}$ defined in Eq.\ \eqref{Precond}. 
Due to the absence of coupling between different ADOs, investigating each row individually is sufficient. 
For first tier ADOs, it holds that
\begin{align} \label{SylvesterPrecond}
  \tilde{\rho}^{(1)}_{Z,j_1} = & (-i \tilde{H}^{\phantom{\dagger}}_{\rm S}  - \gamma^{\phantom{\dagger}}_{j_1})\tilde{\rho}^{(1)}_{Y,j_1} + \tilde{\rho}^{(1)}_{Y, j_1} i\tilde{H}^{\phantom{\dagger}}_{\rm S},
\end{align}
and similar equations are valid for higher tier ADOs.
Matrix equations of this type correspond to the so-called Sylvester matrix equations.\cite{SylvesterOrig} 
Since the two coefficient matrices $-i\tilde{H}_{\rm S} - \gamma_{j_1}$ and $i\tilde{H}_{\rm S}$ have no common eigenvalues, it is guaranteed that a unique solution $\tilde{\rho}^{(1)}_{ Y,j_1}$ exists.\cite{bickley1960matrix} 
Often, the Bartels-Stewart algorithm \cite{101145361573361582} is used to numerically solve Sylvester matrix equations, which requires transformations that convert the two coefficient matrices into upper triangular forms (Schur decomposition).
Since the HEOM in Eq.\ \eqref{HEOMeq} is written in the eigenbasis of the molecular system Hamiltonian, both coefficient matrices have a diagonal Schur form and consequently no further matrix decompositions or factorizations are necessary. 
Overall, a simple backward substitution process solves for the first tier ADO in Eq.\ \eqref{SylvesterPrecond}, resulting in the inexpensive application of the proposed preconditioner. 
The described numerical strategy can also be used to compute the higher tier ADOs with slightly different coefficient matrices.

\subsection{Hierarchy Truncation in the Reduced System Hilbert Space} \label{Trunc}
To improve the efficiency of the HEOM formalism, various methods for the hierarchy termination have been introduced, e.g., closure relations that are local \cite{doi10106311850899} or nonlocal \cite{YAN2004216} in the time domain. Hou $et$ $al.$ extended the latter approach\cite{doi10106314914514} to decrease the computational resources required for steady-state calculations.
Besides neglecting the coupling to $(n+1)$-th and higher tiers, the underlying idea is to set the time derivative of the $n$th tier ADOs to zero. 
This accounts for the influence of the $n$th tier without explicitly setting up its corresponding hierarchical equations. 
Due to the resulting improved computational efficiency \cite{doi10106314914514,PhysRevB97235429}, the combination of this hierarchy truncation scheme with the suggested iterative approach would be the method of choice.

The hierarchy truncation scheme as described in Ref.\ \citenum{doi10106314914514} is defined in the reduced system Liouville space and involves the computation of the inverse of the Liouville operator $\mathcal{L}_{\rm S} O = [H_{\rm S},O]$. 
This numerical operation generally cannot be performed in the reduced system Hilbert space, preventing its straightforward combination with the iterative approach. 
We circumvent this issue by mapping the matrix inversion to a corresponding Sylvester matrix equation as discussed in the following. 
We begin by considering the HEOM in Eq.\ \eqref{HEOM_ss} for $n$th tier ADOs while disregarding the coupling to higher tiers as follows:
\begin{align}
 0 = & -i [\tilde{H}^{\phantom{\dagger}}_{\rm S},\tilde{\rho}^{(n)}_{j_n \ldots j_1}] - \sum^n_{k=1}\gamma^{\phantom{\dagger}}_{j_k}\tilde{\rho}^{(n)}_{j_n \ldots j_1} \nonumber \\
  & -i \sum^{n}_{k=1} \mathcal{\tilde{C}}^{\phantom{\dagger}}_{j_k} \tilde{\rho}^{(n-1)}_{j_n \ldots j_1/j_k}.
\end{align}
Rearranging the equation in terms of a Sylvester matrix equation leads to
\begin{align}
 i \sum^{n}_{k=1} \mathcal{\tilde{C}}^{\phantom{\dagger}}_{j_k} \tilde{\rho}^{(n-1)}_{j_n \ldots j_1/j_k} = & \bigl(-i\tilde{H}^{\phantom{\dagger}}_{\rm S} - \sum^n_{k=1}\gamma^{\phantom{\dagger}}_{j_k} \bigr)\tilde{\rho}^{(n)}_{j_n \ldots j_1} \nonumber \\
  &+ \tilde{\rho}^{(n)}_{j_n \ldots j_1} i\tilde{H}^{\phantom{\dagger}}_{\rm S}.
\end{align}
This result is similar to the equation obtained in previous investigations (see Eq.\ \eqref{SylvesterPrecond}). 
Hence, the same backward substitution process can be utilized to solve for $n$th tier ADOs, and inserting it into the HEOM for $(n-1)$-th tier ADOs leads to the inclusion of $n$th tier without explicitly setting up their hierarchy.

\section{Illustrative Applications} \label{Results}
In this section, we study the performance of the iterative steady-state solver introduced above, which consists of the CGNE method in combination with the preconditioner and the hierarchy truncation scheme (all defined in \cref{Iterative}).
In addition, we compare the required computational resources to methods previously applied for steady-state calculations.
To this end, we employ a model for a molecular junction, where the molecule is described by the following system Hamiltonian:
\begin{align} \label{SystemHamiltonian}
  H_{\rm S}  = & \sum_{m=1}^{N_{\rm el}} \epsilon^{\phantom{\dagger}}_m d^{\dagger}_m d^{\phantom{\dagger}}_m + \sum_{m \neq n} t^{\phantom{\dagger}}_{mn} d^{\dagger}_m d^{\phantom{\dagger}}_n \nonumber \\
  & +\sum_{m > n} U^{\phantom{\dagger}}_{mn} d^{\dagger}_m d^{\phantom{\dagger}}_m d^{\dagger}_n d^{\phantom{\dagger}}_n \nonumber \\
  & + \sum_{\alpha=1}^{N_{\rm vib}} \Omega^{\phantom{\dagger}}_{\alpha} a^{\dagger}_{\alpha} a^{\phantom{\dagger}}_{\alpha} + \sum_{\alpha,m,n} \lambda^{\alpha}_{mn} (a^{\phantom{\dagger}}_{\alpha} + a^{\dagger}_{\alpha}) d^{\dagger}_m d^{\phantom{\dagger}}_n .
\end{align}
Here, the parameter $\epsilon_m$ denotes the energy of the molecular electronic state $m$ and $d^{\phantom{\dagger}}_m/d^{\dagger}_m$ is the corresponding annihilation/creation operator. 
Hopping between different sites is governed by the parameter $t_{mn}$ and $U_{mn}$ characterizes the Coulomb interaction strength.
The vibrational degrees of freedom are modeled as harmonic oscillators with frequencies $\Omega_{\alpha}$, where $a^{\phantom{\dagger}}_{\alpha}/a^{\dagger}_{\alpha}$ is the lowering/raising operator of vibrational mode $\alpha$ and the electronic-vibrational coupling strength is described by the parameter $\lambda^{\alpha}_{mn}$. 
We refer to a model where $\lambda^{\alpha}_{mn}=0$ and $U_{mn}=0$ as noninteracting. 
In the following, we consider a symmetric drop of the bias voltage $\Phi$ at the molecule-lead contacts, i.e., the chemical potentials in the left and right electrode are given by $\mu_L=\Phi/2$ and $\mu_R=-\Phi/2$, respectively. 
In total, the molecule consists of $N_{\rm el}$ electronic sites and $N_{\rm vib}$ vibrational degrees of freedom. 
Considering a set of $\nu_{\alpha}$ vibrational basis states for mode $\alpha$, the dimension of the associated reduced system Hilbert space is $N_{\rm S} = 2^{N_{\rm el}} \Pi^{N_{\rm vib}}_{\alpha=1} \nu_{\alpha}$.

In the following, we test the performance of the iterative approach to solve the HEOM with the system Hamiltonian, Eq.\ \eqref{SystemHamiltonian}, and compare it to other numerical strategies. 
For comparison, we have selected two reference methods often used in this context, including direct solving \cite{PhysRevB94201407,PhysRevB97235429} and a real-time propagation algorithm \cite{doi10106314914514}. 
The first reference approach uses the LU decomposition in combination with the time nonlocal truncation scheme introduced in Ref.\ \citenum{doi10106314914514}. 
The second reference approach includes Runge-Kutta-Fehlberg 4(5) adaptive time step-size integration \cite{doi101021ct3003833} and utilizes sparse matrix-vector multiplication. 
To improve the computational efficiency, we have implemented this method to take advantage of the sparsity feature of the ADOs.\cite{doi10106314914514} 
Compared to direct solving, we refrain from utilizing the aforementioned hierarchy truncation scheme since it was reported to increase the computational resources required by real-time propagation algorithms \cite{doi10106314914514}. 
The initial conditions for both the iterative and real-time propagation approach are set to the state where the electronic states of molecule are unoccupied. 
All simulations were run with standard optimization flags using a single core of a node with two Intel(R) Xeon(R) E6252 Gold central processing units.

To cover a broad range of parameters, we first investigate a noninteracting model system and proceed with a study of two interacting cases. 
Finally, we examine the efficiency of the improvements introduced in the previous section.

\subsection{Noninteracting System with Multiple Electronic Sites} \label{NMES}
We first consider a model system consisting of $N_{\rm el}$ electronic states, which are serially coupled to form a linear chain and are labeled as $1,\ldots,N_{\rm el}$, starting with the leftmost state. 
In this scenario, the purely electronic hopping terms are zero except between neighboring sites, i.e., $t_{mn}= t( \delta_{m+1,n} + \delta_{m,n+1})$. 
The left (right) end of the chain is coupled to the left (right) electrode with the coupling matrix element $V_{L,m}=V\delta_{m,1}$ ($V_{R,m}=V\delta_{m,N_{\rm el}}$). 
We assume the on-site energies to be $\epsilon_m = \SI{0.3}{\electronvolt}$ and set the temperature of the electrodes to $T=\SI{200}{\kelvin}$. 
For this model system, it is sufficient for converged results to employ second tier calculations while including $N_{\rm P}=13$ Pad\'{e} poles. 
In total, $N_{\rm ADO}=1596$ ADOs ($N_{\rm ADO}=28$ ADOs when using the hierarchy truncation scheme) need to be taken into account.

\begin{figure}
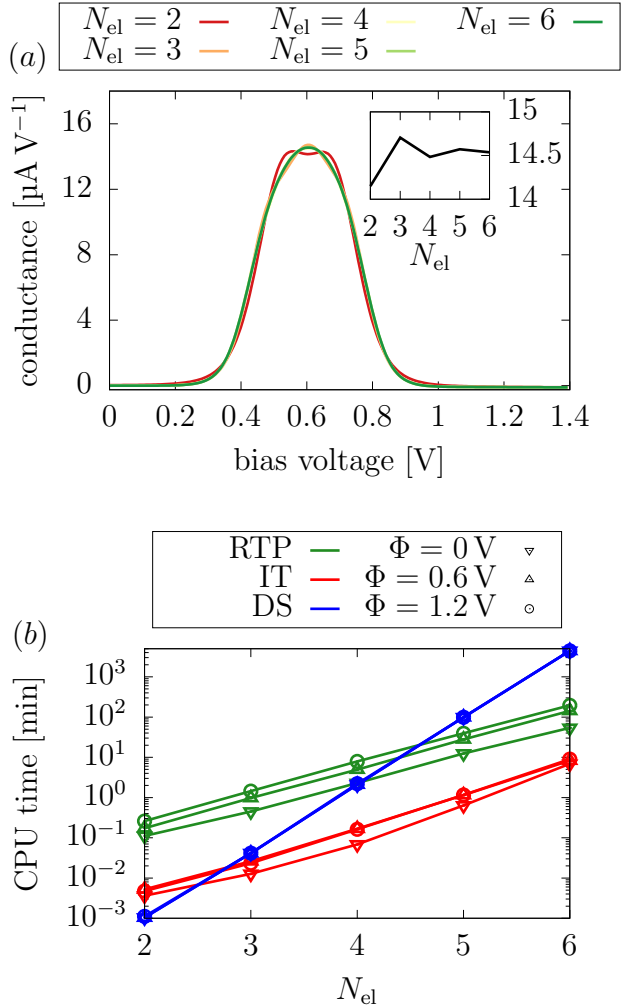

 \input{Cond.tex}
 
 \vspace{0.5cm}
 
 \input{TimeMES.tex}
 \caption{(a) Conductance as a function of the applied bias voltage. The inset shows the length dependency of the conductance at a bias voltage $\Phi=\SI{0.6}{\volt}$. (b) Comparison of the required CPU time for steady-state calculations. The different colors correspond to the different methods, while the symbols represent the considered bias voltages. The following parameters were used: $W_{\rm L/R}=\SI{5}{\electronvolt}$, $\Gamma=2 \pi V^{2} =\SI{0.05}{\electronvolt}$ and $t= \Gamma$.}
    \label{ResultsChain}
\end{figure}

This model has been studied before (see Refs.\ \citenum{zeng2002delocalization,kim2002even,asai2005theory,PhysRevB83075437,taranko2012transient}) and interesting phenomena such as delocalization \cite{zeng2002delocalization} or even-odd parity effects in the shot noise\cite{kim2002even} were reported. 
\Cref{ResultsChain}(a) depicts the conductance as a function of the bias voltage for a maximal chain length $N_{\rm el}=6$. 
The conductance exhibits a step located around $\Phi=\SI{0.6}{\volt}$ that corresponds to the onset of resonant transport of electrons through the electronic sites of the serial chain.
For the bias voltage $\Phi=\SI{0.6}{\volt}$, the inset shows the even-odd parity effects of the conductance, which decrease for increasing chain length in this parameter regime.
The conductance is negative for large bias voltages due to the finite band-width. 
More detailed investigations can be found in Ref.\ \citenum{asai2005theory}.

To analyze the performance of the methods, \cref{ResultsChain}(b) shows the required CPU time of steady-state calculations for increasing chain lengths.
The results of the iterative scheme and the two reference methods are depicted for three different bias voltages. 
We refer to the real-time propagation algorithm as RTP, the iterative approach as IT and the direct solving method as DS and begin with examining the differences and similarities between the latter two strategies.
The required CPU time of the method DS is independent of any model parameter and hence, remains constant for all applied bias voltages. 
In contrast to this, iteratively solving for the steady-state reveals a slight bias dependency. 
In this parameter regime, the unoccupied molecule represents a better initial state for low bias voltages; hence, less iterative steps and, consequently, a reduced amount of CPU time are required.
The crucial difference between the two algorithms is the scaling with respect to the considered chain length, i.e., the dimension of the reduced system Hilbert space. 
The numerical effort of method DS is mainly determined by two routines. 
On the one hand, it involves the LU factorization of a matrix with dimension $N_{\rm ADO} \cdot 2^{2 N_{\rm el}}$.
On the other hand, the hierarchy truncation scheme requires the multiplication of matrices with a size $2^{2N_{\rm el}}$.
Since both numerical operations scale with the third power of the corresponding matrix size, the CPU time of method DS is roughly proportional to $2^{6 N_{\rm el}}$.
This is in contrast to method IT whose underlying CGNE algorithm, preconditioner, and the hierarchy truncation scheme only incorporate multiplications of matrices defined in the reduced system Hilbert space. 
Hence, the CPU time of this approach roughly scales with $2^{3 N_{\rm el}}$, explaining the enhanced computational efficiency with respect to increasing chain sizes in comparison to that of DS.

Next, we examine the differences between the methods IT and RTP.
Compared to that of the iterative procedure, the CPU time required by the real-time propagation approach shows an enhanced dependency on the applied bias voltage.
For the investigated serially coupled linear chain, the required computational resources rise along with the increasing bias voltage $\Phi$. 
This is due to the fact that the unoccupied molecule is a more suitable initial state for propagation in the nonresonant low-voltage regime.
In contrast to the schemes IT and DS, the CPU time required for steady-state calculations with RTP does not follow a certain scaling rule. 
In this method, the numerically most costly operation is sparse matrix-vector multiplication. 
Hence, the numerical complexity is dictated by the total amount of nonzero entries, which is determined by the specific Hamiltonian that describes the single-molecule junction as well as the off-diagonal elements of the level-width functions $\Gamma_{Kmn}(\epsilon)$ (see Ref.\ \citenum{doi10106314914514}). 
Nevertheless, the results in \cref{ResultsChain}(b) indicate that the scaling is comparable to that of the proposed iterative algorithm, i.e., roughly proportional to $2^{3 N_{\rm el}}$.
Hence, the main advantage of the iterative method over the real-time propagation algorithm is the reduced amount of required CPU time, which results from a decreased number of steps.

\subsection{Electronic Transport with Coulomb Interactions} \label{IN}
Next, we extend our studies to an interacting system by incorporating Coulomb interactions into the model of the previously studied serially coupled linear chain. 
We consider a maximal chain length of $N_{\rm el}=3$ and set the Coulomb interaction strength to $U_{mn}=\SI{0.05}{\electronvolt}$; all other parameters are unchanged. 
In this scenario, the hierarchy does not automatically terminate after the second tier.
Test calculations show that including third tier contributions provides converged results for the parameters considered.

\begin{figure}
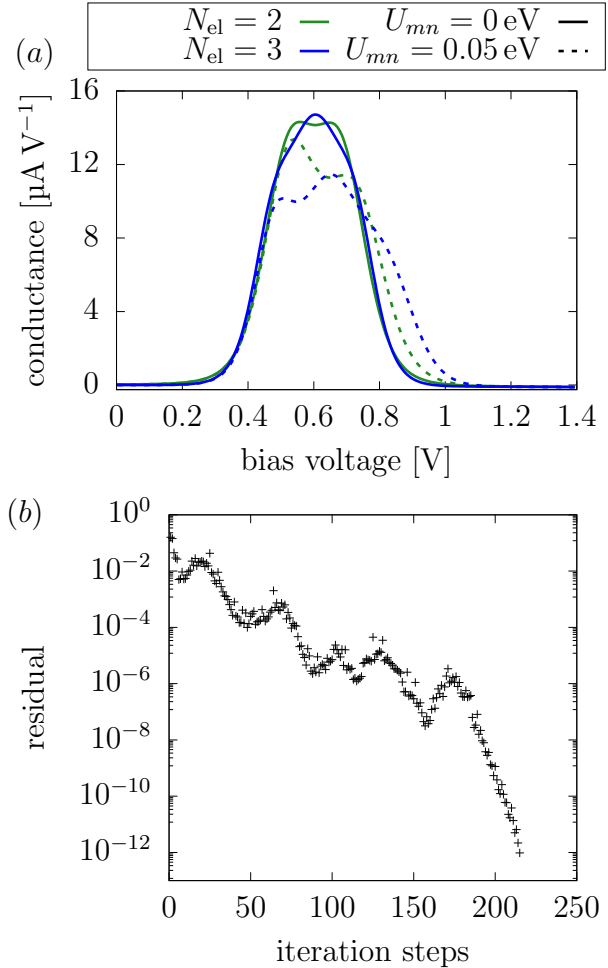

 \input{UChain.tex}
 \input{Res.tex}
 \caption{(a) Conductance as a function of the bias voltage of a serially coupled linear chain with and without Coulomb interactions. The maximal investigated chain length is $N_{\rm el}=3$. (b) The residual (Eq.\ \eqref{residual}) as a function of the required iterative steps for a chain length $N_{\rm el}=2$. We used the following parameters: $W_{\rm L/R}=\SI{5}{\electronvolt}$, $\Gamma=2 \pi V^{2} =\SI{0.05}{\electronvolt}$ and $t= \Gamma$.}
 \label{ResultsUChain}
\end{figure}

The corresponding conductance-voltage characteristic is shown in \cref{ResultsUChain}(a).
Due to the Coulomb interaction, doubly occupying the molecular electronic states requires an additional charging energy. 
This results in a shift of some of the transport channels to larger bias voltages, which explains the difference between the conductances of $U_{mn}=\SI{0.05}{\electronvolt}$ and $U_{mn}=0$.
For a further analysis, see Ref.\ \citenum{asai2005theory}.

\Cref{TResultsUChain} compares the numerical resources required by the proposed iterative method and the two reference approaches as a function of the chain length.
\begin{table*}[tb]
  \begin{center}
 \begin{tabular}{l|c|c|c|c|c|c}
 \hline\hline
         & \multicolumn{3}{c|}{$N_{\rm el}=2$} & \multicolumn{3}{c}{$N_{\rm el}=3$} \\
         \hline
  method & $\SI{0}{\volt}$ & $\SI{0.6}{\volt}$ & $\SI{1.2}{\volt}$ & $\SI{0}{\volt}$ & $\SI{0.6}{\volt}$ & $\SI{1.2}{\volt}$ \\
 \hline
 IT    & $0.43$ & $0.74$ & $0.32$ & $7.21$  & $8.34$  & $2.54$  \\
 RTP   & $4.45$ & $7.03$ & $10.4$ & $59.8$  & $42.4$  & $22.5$  \\
 DS    & $55.9$ & $55.9$ & $55.9$ & $13600$ & $13600$ & $13600$ \\
 \hline \hline
 \end{tabular}
\end{center}
 \caption{Comparison of the Required CPU Time in Minutes for Three Steady-State Methods.
 The three columns correspond to the three different bias voltages, $\Phi=0$, $\Phi=\SI{0.6}{\volt}$ and $\Phi=\SI{1.2}{\volt}$}
 \label{TResultsUChain}
\end{table*}
Setting the hierarchy level to three leads to a larger number of ADOs ($N_{\rm ADO}=45500$, $N_{\rm ADO}=1596$ when the hierarchy truncation scheme is used), resulting in a significant increase of the required computational resources compared to those of the previously investigated second tier calculations in the noninteracting model. 
In general, the  CPU times of the methods RTP and IT show an enhanced bias voltage dependency. 
As discussed in \cref{GCME}, the parameter that mainly determines the convergence rate is the squared condition number of the matrix which represents the hierarchical equations in the Liouville space.
In this model system, the condition number has the largest value for bias voltage $\Phi=\SI{0.6}{\volt}$ and lower values for $\Phi=0$ and $\Phi=\SI{1.2}{\volt}$, respectively.
Besides this, the results can be rationalized as in the case without Coulomb interactions. 
Overall, iteratively solving for the steady-state is the most efficient method.

To further illustrate the numerical performance of the proposed iterative method, we show in \cref{ResultsUChain}(b) the convergence behavior for a chain length of $N_{\rm el}=2$ at the bias voltage $\Phi=\SI{1.2}{\volt}$.
The depicted residual is given by \cite{ZHANG20119380}
\begin{align}
 R = || \mathcal{M}(X)||_{\text{F}}, \label{residual}
\end{align}
with $X$ and $\mathcal{M}(X)$ defined in \cref{GCME} and the Frobenius norm $||.||_{\text{F}}$.
Characteristic for the CGNE method is the nonmonotonous decay of the residual.
Converged results are obtained for the threshold $R\le 10^{-12}$.

\subsection{Vibrationally Coupled Charge Transport}
To further demonstrate the efficiency of the iterative approach, we investigate a representative model single-molecule junction, which incorporates the coupling between electronic and vibrational degrees of freedom. 
Specifically, we consider two electronic states with an energy of $\epsilon_{m} = \SI{0.05}{\electronvolt}$ and include a single harmonic oscillator with a frequency of $\Omega_{1} = \SI{0.1}{\electronvolt}$. 
The interstate coupling is set to $t=\SI{0.15}{\electronvolt}$ and the electronic-vibrational coupling strength is set to $\lambda^{1}_{mn} = \delta_{mn} \SI{0.05}{\electronvolt}$. We assume all coupling matrix elements to be equal, i.e., $V_{K,m}=V$.
In order to obtain converged results, $N_{\rm P}=13$ Pad\'{e} poles and second tier calculations are required at a temperature of $T=\SI{100}{\kelvin}$. 
Recently, similar model systems were investigated in Refs.\ \citenum{PhysRevB95085423,PhysRevB87075422,PhysRevB67125323,PhysRevB81205408,PhysRevB90125450}.

\begin{figure}
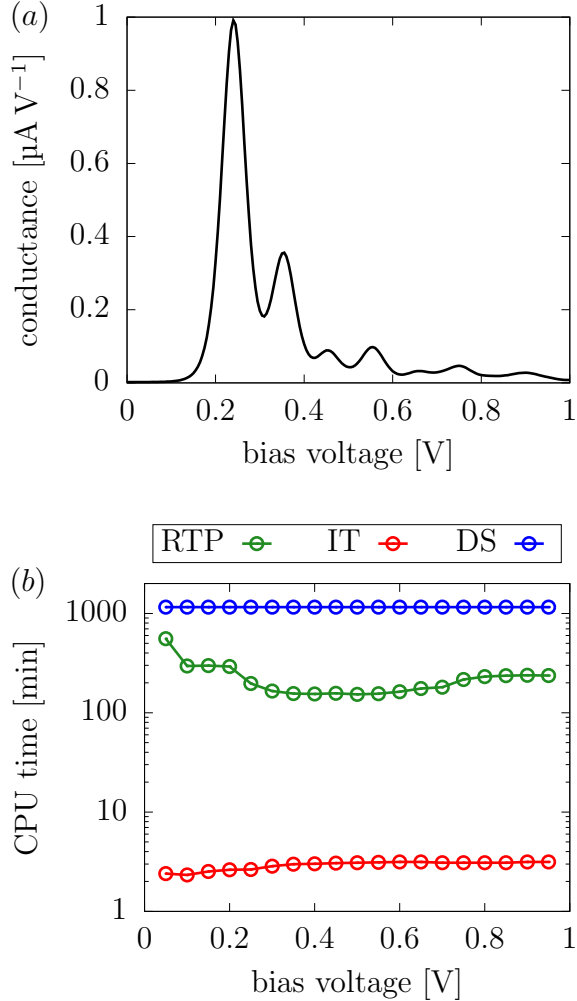

 \input{VIB.tex}

 \vspace{0.5cm}

 \input{TimeVIB.tex}
 \caption{(a) Conductance as a function of the bias voltage of a single-molecule junction incorporating electronic-vibrational coupling. 
 (b) Required CPU time for steady-state calculations. 
 The different colors correspond to the different methods. 
 The following parameters were used: $W_{\rm L/R}=\SI{5}{\electronvolt}$ and $\Gamma=2 \pi V^{2} =\SI{0.001}{\electronvolt}$.}
 \label{ResultsVIB}
\end{figure}

The conductance-voltage characteristic of the model single-molecule junction is depicted in \cref{ResultsVIB}$(a)$.
It exhibits multiple peaks that result from the onset of resonant transport processes. 
In addition to elastic transport, the coupling between the electronic and vibrational degrees of freedom triggers a number of inelastic transport channels, which include the excitation of multiple vibrational quanta.\cite{PhysRevB70125406,Galperin_2007}

The CPU times required for steady-state calculations with the iterative scheme and the two reference methods are depicted in \cref{ResultsVIB}$(b)$. 
The algorithm RTP shows an enhanced dependency on the applied bias voltage in comparison to the previous investigations. 
The increasing CPU time for low bias voltages is attributed to the occurrence of long-time oscillations in the current and populations that originate from the inclusion of electronic-vibrational coupling. 
Similarly, the iterative method IT requires more computational resources to solve for the steady-state in the scenario of high bias voltages. 
The condition number increases for high bias voltages requiring longer CPU times for iterative steady-state calculations. 
In contrast, the consumed computational resources of the approach DS are independent of the applied bias voltage. 
Including electronic-vibrational coupling, the dimension of the reduced system Hilbert space scales with $2^{N_{\rm el}} \nu_{1} $, with $N_{\rm el}=2$ and $\nu_{1}=11$.
As shown in \cref{NMES}, the approach IT is capable of handling large system sizes particularly well and shows the best overall efficiency.

\subsection{Performance of Preconditioner and Hierarchy Truncation Scheme}
In the following, we illustrate the performance of the preconditioner and the hierarchy truncation scheme in the reduced Hilbert space introduced in \cref{PrecondSec,Trunc}. 
They significantly improve the computational efficiency when iteratively solving for the steady-state.
To show this, we introduce three additional variants of the proposed steady-state solver.
We refer to the iterative approach without truncating the hierarchy as IT/T and the procedure IT/P does not include the preconditioner.
The method IT/PT incorporates neither of the two strategies.

As a benchmark model, we use the interacting system of the serially coupled linear chain investigated in \cref{IN} for two electronic sites.
\begin{table*}[tb]
\begin{center}
 \begin{tabular}{l|c|c|c|c|c|c|c}
 \hline \hline
  \multicolumn{2}{c|}{} & \multicolumn{3}{c|}{CPU time [min]} & \multicolumn{3}{c}{\# steps} \\
  \hline
  method & Ram [Mbyte] & $\SI{0}{\volt}$ & $\SI{0.6}{\volt}$ & $\SI{1.2}{\volt}$ & $\SI{0}{\volt}$ & $\SI{0.6}{\volt}$ & $\SI{1.2}{\volt}$  \\
 \hline
 IT/PT & $57.2$  & $26.7$ & $21.5$ & $18.3$ & $9454$ & $8538$ & $7685$ \\
 IT/T  & $57.2$  & $1.2$  & $0.92$ & $0.31$ & $1565$ & $1389$ & $460$ \\
 IT/P  & $1.01$  & $2.52$ & $1.71$ & $0.64$ & $2301$ & $1569$ & $591$ \\
 IT    & $1.01$  & $0.43$ & $0.74$ & $0.32$ & $271$  & $493$  & $215$  \\
 RTP   & $69.6$  & $4.45$ & $7.03$ & $10.4$ & & &\\
 DS    & $38900$ & $55.9$ & $55.9$ & $55.9$   & & &\\
 \hline \hline
 \end{tabular}
\end{center}
 \caption{Required Computational Resources of Different Steady-State Approaches. The three columns correspond to the three different bias voltages, $\Phi=0$, $\Phi=\SI{0.6}{\volt}$ and $\Phi=\SI{1.2}{\volt}$.
 The investigated system corresponds to an interacting model system with a chain length of $N_{\rm el}=2$. 
 The parameters are chosen as in \cref{IN}}\label{ResultsIntro}
 \end{table*}
The CPU time, working memory, and number of iterative steps required by the different variants are summarized in \cref{ResultsIntro}.
For comparison, we added the corresponding information for the two reference methods.

The results show that the IT/PT approach requires, for all bias voltages, a large amount of iterative steps, which indicates the large value of the condition number.
In order to increase the efficiency, applying a preconditioner is inevitable. 
The approach IT/T incorporates the preconditioner introduced in \cref{PrecondSec}.
Even though it belongs to a simpler form of preconditioning, it significantly reduces the number of iterative steps and thus the CPU time. 
Since no additional variables are necessary, its application has no impact on the memory usage.

Instead of utilizing a preconditioner, the method IT/P uses the hierarchy truncation scheme in the reduced Hilbert space. 
It improves the computational efficiency because it avoids setting up and storing the third tier hierarchy, resulting in a reduced number of ADOs ($N_{\rm ADO}=1596$). 
As a result, less working memory is used, and both the number of iterative steps and the CPU time decrease, since applying the hierarchy truncation scheme lowers the condition number.
Originally, Hou $et$ $al.$ introduced this scheme for real-time propagation methods.\cite{doi10106314914514}
In contrast to our numerical findings for steady-state computations, the authors report increasing CPU times compared to those of the standard HEOM implementation.
While less iterative steps are needed for the proposed iterative approach, it does not necessarily reduce the number of real-time propagation steps.

The combination of both improvements results in a particularly efficient iterative steady-state solver for the HEOM method compared to either direct solving or real-time propagation algorithms. 
Furthermore, the dependency of the bias voltage on the CPU time is minimized.

\section{Conclusion} \label{Conclusion}
In this paper, we have introduced an iterative steady-state solver for the HEOM formalism by mapping the hierarchically coupled equations onto general coupled matrix equations. 
This formulation enables all numerical operations to be performed in the reduced system Hilbert space, resulting in both a superior memory usage and less time-consuming computations compared to those for previously applied steady-state approaches.
As the underlying iterative algorithm, a conjugate gradient method \cite{ZHANG20119380} is utilized. 
We employ an efficient and inexpensive preconditioning technique defined in the reduced system Hilbert space, which is inevitable for iterative methods.
To further improve the efficiency, we have extended the hierarchy truncation scheme introduced by Hou $et$ $al.$ \cite{doi10106314914514} to the reduced system Hilbert space. 
The performance of the iterative solving technique was demonstrated by applications to different models of charge transport in molecular junctions. 
The results demonstrate the efficiency of the method, particularly with respect to the scaling of the numerical effort for increasing dimensions of the reduced system Hilbert space.

While we have focused on nonequilibrium charge transport in the present work, other interesting processes such as heat transport through single-molecule junctions\cite{doi101146annurevphyschem040215112103,doi10106314928192,doi10106315075620} can also be investigated with the iterative steady-state solver.
Furthermore, the extension of the method to systems that additionally include bosonic environments is possible\cite{batge2021nonequilibrium}. 
Because of the small amount of memory available on current GPGPU generations, such investigations have been restricted to small model systems\cite{Kreisbeck2011,Kreisbeck2012,Hein2012,Kreisbeck2014,doi101021acsjctc5b00488,C6FD00088F}. 
Being efficient in terms of memory usage, the implementation of the proposed iterative method on GPGPU devices is thus very promising.

\section*{Acknowledgements}
This paper is dedicated to Yoshitaka Tanimura on the occasion of his 60th birthday. 
MT thanks Yoshitaka Tanimura for insightful discussions on nonequilibrium quantum dynamics and the HEOM method. 
We thank T. Novotn\'{y} for introducing us to the concept of Sylvester matrix equations and J. B{\"a}tge for inspiring discussions.
This work was supported by a research grant of the German Research Foundation (DFG). 
Furthermore, support from the state of Baden-{W\"urttemberg} through bwHPC and the DFG through Grant INST 40/575-1 FUGG (JUSTUS 2 cluster) are gratefully acknowledged.

\bibliography{./Bib/paper.bib}
\end{document}